\documentclass[twocolumn,prb,aps,floatfix,superscriptaddress]{revtex4-1}
\usepackage{amsmath,amssymb,bm,epsfig,color}
\usepackage{siunitx}

\renewcommand{\vec}[1]{\boldsymbol{#1}}

\begin{document}

%\title{Atom--Graphene van der Waals  Interaction: Strain and Correlation Effects}
%\title{Designing van der Waals  Interactions by Straining Graphene with Atoms on Top}
\title{Adsorption by design: tuning atom-graphene van der Waals interactions via
mechanical strain}
\author{Nathan S. Nichols} 
\affiliation{Department of Physics, University of  Vermont, Burlington, VT 05405}
\author{Adrian Del Maestro} 
\affiliation{Department of Physics, University of  Vermont, Burlington, VT 05405}
\author{Carlos Wexler} 
\affiliation{Department of Physics and Astronomy, University of Missouri, 
Columbia, MO 65211}
\author{Valeri N. Kotov}
\affiliation{Department of Physics, University of  Vermont, Burlington, VT 05405}

\date{\today}
\begin{abstract}

We aim to understand how the van der Waals force between neutral  adatoms and a
graphene layer is modified by uniaxial strain and electron correlation effects.
A detailed analysis is presented for three atoms (He, H, and Na) and graphene
strain ranging from weak to moderately strong. We show that the van der Waals
potential can be significantly enhanced by strain, and present applications of
our results to the problem of elastic scattering of atoms from graphene. In
particular we find that  quantum reflection can be significantly suppressed by
strain, meaning that dissipative inelastic effects near the surface  become of
increased importance.  Furthermore we introduce a method to independently
estimate the Lennard-Jones parameters used in an effective model of He
interacting with graphene, and determine how they depend on strain.  At short
distances, we find that strain tends to reduce the interaction strength by
pushing the location of the adsorption potential minima to higher distances
above the deformed graphene sheet.  This opens up the exciting possibility of
mechanically engineering an adsorption potential, with implications for the
formation and observation of anisotropic low dimensional superfluid phases.

\end{abstract}
\pacs{ }

\maketitle

\section{Introduction}
\label{sec:intro}

van der Waals (vdW) or dispersion forces play an especially important role at
interfaces involving atomically thin materials, such as graphene and
structurally  similar materials, including transition-metal dichalcogenides
(\emph{e.g.} MoS$_{2}$).  These can form the building blocks of the so-called
van der Waals heterostructures \cite{vd1}.  vdW interactions are fundamentally
and practically important,  as they   reflect  the polarization properties of
materials and are sensitive to Coulomb interactions. In addition, as will be
discussed below, they can depend strongly on material deformations, both through
modifications of the electronic structure which affects the polarization, and
the changes induced in the electron-electron interactions. 

Two-dimensional materials can withstand large strains without rupture, offering
unique opportunity for exploration of large strains.  In graphene,
uniaxial strain effects (most notably along the ``armchair" or the ``zig-zag"
directions)  have been studied theoretically within the non-interacting
tight-binding framework \cite{pereira,ribeiro,choi,pellegrino,pellegrino1}.
This  theoretical work was mostly motivated by experimental investigations of
graphene's mechanical properties: graphene was confirmed to be the strongest
material ever measured \cite{lee}, and is able to sustain reversible elastic
(uniaxial) strain of $\delta \approx 20\%$ \cite{ni,kim}.  In addition, strain
and ripple formation can coexist and affect the functionalization properties of
graphene, such as the adsorption of atomic hydrogen ()which quickly turns graphene
into an insulator) \cite{elias}.  One can also imagine  many possibilities for
local strain engineering, including the creation of strain profiles that can produce
desired electronic properties, such as confinement, and surface states.
\cite{pereira1}.  Strain plays an important role in the electronic structure of
numerous two-dimensional (2D) materials as described in a recent review
(Ref.~[\onlinecite{Maria}]), and general strain configurations
corresponding to gauge fields with different symmetries have to be taken into
account.  From now on we will consider only uniaxial strain, as it is one of the
simplest deformations and is amenable to a practically complete
theoretical analysis of vdW forces in terms of their strain and correlation
dependence.

%
% ------------------------------------------------------------------------------- 
\begin{figure}[t]
\begin{center}
\includegraphics[width=1.0\columnwidth]{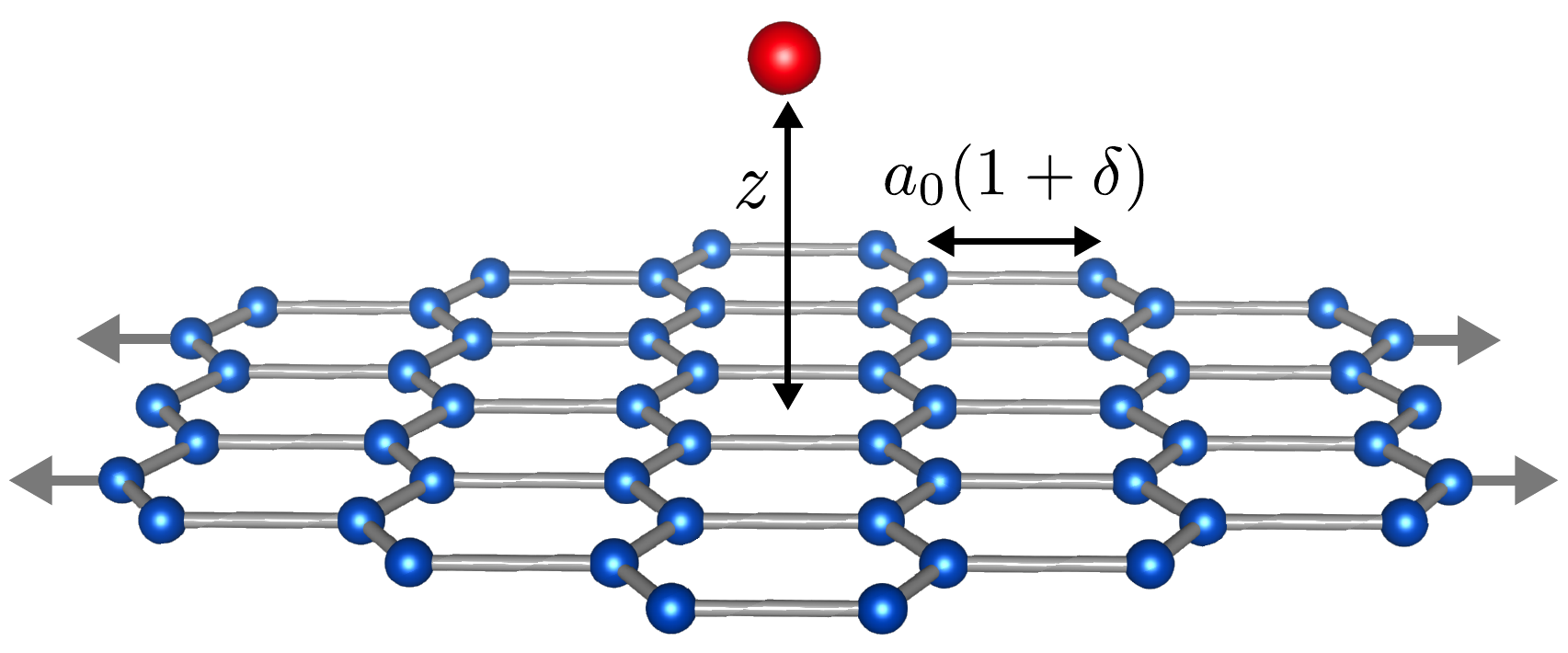}
\end{center}
\caption{(Color online) An adatom located a distance $z$ above a graphene sheet subject
to mechanical strain ($\delta$) along the indicated armchair direction.}
\label{fig:strain}
\end{figure}
% ------------------------------------------------------------------------------- 
%

The electronic structure  under uniaxial strain shows strong directional
dependence -- in particular, the armchair deformation shown in
Fig.~\ref{fig:strain}, results in a tendency towards the system becoming more
one dimensional, while a zig-zag stress leads to dimer formation beyond a
critical value $\delta_{c} \approx 23\%$ which generates a gap (via a
topological Lifshitz transition) in the electronic spectrum
\cite{pereira,ribeiro}.  For weak strain, the electronic spectrum is
anisotropic (elliptical, with different Fermi velocities $v_y \neq v_x$) in
both cases.  This behavior creates a rich variety of possibilities for
interplay between strain-induced polarization and electron-electron
interactions and is our subject of interest.

Other examples of graphene-based lattices with   anisotropic Dirac
excitations include: (1)  graphene superlattices
\cite{chpark,chpark2,rusponi,herb}, (2) tunable honeycomb optical lattices
\cite{tarruell}, and (3) molecular graphene, formed by manipulation of carbon
monoxide molecules over conventional 2D electron systems \cite{gomes}.  For
example,  a high anisotropy (ratio of Fermi velocities $v_y/v_x \approx 0.5$) has
been achieved in epitaxial graphene modulated on an island superlattice
\cite{rusponi}.  These recently developed systems provide further opportunities
for manipulation and tuning of the conventional graphene electronic structure
and thus exploration of the anisotropy-related effects and their
consequences for vdW forces.

van der Waals forces between graphene sheets (at distance $d$, large enough to
eliminate direct hopping between layers), have been a subject of considerable
attention \cite{vd1, vd2, vdw2,vd4,vd5,vd6,vd7,vd8,vdw1,vd10,vd11,vd12}. For
Dirac systems in 2D, in particular graphene, the force decreases as fourth power
of the distance, $|F_{vdW}(d)|= C_{vdW}/d^4$, and is fairly weak compared to
relativistic systems (due to the small value of the Fermi velocity compared to
the speed of light). A fundamental and practical question arises: \emph{Can
this force be enhanced?}

In a recent work \cite{vdw-anand}, based on the  random
phase approximation (equivalent to the Lifshitz theory) \cite{ll1,ll2}, 
 we have found that the Dirac anisotropy (i.e.
strain)  can substantially enhance the force resulting from the growth of the
polarization with increased anisotropy. 
Experimentally realizable values of strain show 10 times increases of the force.  Moreover, as emphasized in recent works \cite{vd10, vdw-anand},
the vdW interaction is very sensitive to the  Coulomb  coupling and its 
renormalization; this effect is particularly strong for large strain. Additionally, exchange-correlation phenomena is enhanced in strained graphene, such as the tendency towards itinerant ferromagnetism
\cite{adatom-anand}. The absence of conventional screening of the
Coulomb interactions when the Fermi energy is at (or close to) the Dirac point causes strong electron correlation effects in graphene.
This is typically the case when graphene is produced, e.g. by mechanical
exfoliation \cite{novoselov, novoselov1, geim, geim1, neto}.  The location of
the Fermi energy can also be easily shifted to the Dirac point by application
of backgate voltage, i.e.  due to the strong electric field effect --- one of
the most important characteristics of the material.  A recent overview of
interactions in graphene can be found in Ref. ~[\onlinecite{kotov}].  The linear
spectrum indicates the effective interaction parameter,  typically denoted by $\alpha =
e^2/\hbar v_F$, is doping independent. For suspended graphene, i.e.  without
the dielectric screening from a substrate, the coupling reaches its maximum value of $\alpha
\approx 2.2$.

The main goal of this paper is  to investigate how atoms of different types
interact with uniaxially strained graphene, which we consider as a prototype
strained 2D material.   We perform calculations  for distances up to  70 nm 
and restrict ourselves to  T = 0 since it is well documented 
\cite{vdw3,vdw4,vdw1} that finite temperature effects are negligible in this distance range. 
The study of such atom--2D material interfaces allows
us to explore the effects of strain and interactions within the material in
their most pure form (since interlayer screening of the vdW force
is not present in this case).    While previous works have been devoted to vdW interactions of
atoms with isotropic graphene \cite{vdw3, vdw2,AVDW-2, vdw5, vdw4, AVDW-5},
our work focuses on the effects of strain and correlations. 
One of our principal  results is  that the vdW interaction
increases with strain and the relative magnitude of this effect does not depend strongly
on the type of atom, i.e. on its mass and polarizability.   On the other hand, the vdW interaction is quite sensitive to
graphene's electron-electron interaction coupling constant. 
  Thus, atoms can act as  amplifiers  of
the strain-induced polarization properties of the 2D material, which in turn
can lead to profound consequences for the atomic behavior itself near the
surface. As an application of our theory we calculate the effect of strain on
the quantum reflection amplitude, the probability that a low-energy impinging
atom will be reflected from the surface, and find that it can be dramatically
reduced. 

Increased density of adatoms and the ability to mechanically tune the
van der Waals attraction between them and the graphene sheet opens up the
possibility of investigating low dimensional collective many-body effects near the
surface.  The local anisotropy of the deformed graphene lattice structure
will strongly affect the physics of adsorption.  The search for an ideal and
controllable substrate onto which a light gas (H or He) can be adsorbed to form
a 2D quantum liquid (or superfluid) has been an area of active research for
nearly fifty years.\cite{Bruch:2007pa}  The key requirements for such a
substrate include (1) that it is atomically flat and regular, as disorder would
tend to localize the fluid and (2) it be only weakly polarizable, to prevent
the formation of fully classical wetting layers.  Originally, graphite appeared
to provide an ideal surface in these regards, and its  helium
adsorption phase diagram as a function of density and temperature is well
understood both experimentally \cite{Bretz:1971jo,Zimmerli:1992hz,
Greywall:1993eg, Dash:1994cp, Nakamura:2014vx} and via numerical quantum Monte
Carlo simulations \cite{Abraham:1987ct,Pierce:1999cn,Corboz:2008iq}.  It
includes a commensurate $\sqrt{3} \times \sqrt{3}$~$\mathrm{R}~30^{\circ}$
phase (where helium atoms occupy $1/3$ of the strong binding sites located at
hexagon centers) and possible striped incommensurate and reentrant fluid phases
at high densities, but the first adsorbed layer appears to lack any signatures
of a more exotic quantum liquid.

While experiments are currently lacking, a single sheet of graphene seems to
be an even more appealing substrate for adsorbing quantum fluids as a 10\% reduction in the binding energy for a monolayer of helium (compared to graphite) supresses classical wetting \cite{Gordillo:2009jb}. This
idea was explored via a series of recent zero temperature diffusion Monte Carlo
studies \cite{Gordillo:2011jb, Gordillo:2012fl, Gordillo:2013jc} which reported
the observation of superfluidity in the first layer of helium on graphene and
even the presence of the fleeting \emph{supersolid} phase (where long range
off-diagonal and positional order coexist). These results have proven
controversial, large scale finite temperature grand canonical quantum Monte Carlo
simulations  \cite{Happacher:2013ht} find no evidence of either first layer
superfluid response or supersolidity, with the discrepancy being blamed on
population size bias in diffusion Monte Carlo \cite{Boninsegni:2012ps}.  The
exact nature of adsorbed helium on graphene at low temperature thus remains an
open question.

The most important ingredient in numerical simulations of helium on graphene is
the specific form of the interaction potential between an adsorbate atom and
the graphene sheet and is usually taken as a summation of repulsive hard core
and attractive van der Waals interactions \cite{Carlos:1980cf} which may depend
on a number of phenomenological parameters.  By exploiting our knowledge of the
electronic polarizability of the graphene sheet, we have devised a method that
enables the independent determination of these parameters by fitting the long
distance tail of the van der Waals potential computed within the continuum
limit to predictions from the effective microscopic theory.  This allows us to
investigate both the accuracy of commonly used model parameters for isotropic
graphene as well as the effects of strain on their values.  We find that while
increasing uniaxial strain enhances the long distance van der Waals attraction,
it can have the opposite effect at short distances, leading to an overall
softening of the adsorption potential with exciting consequences for the
energetic feasibility of proximate and possibly anisotropic superfluid phases.
These trends are confirmed via \emph{ab initio} calculations of the interaction
energy between a helium atom and an aromatic molecule composed of 24 carbon
atoms, coronene.

The rest of the paper is organized as follows. In Section~\ref{sec:atom_graphene}
we describe our results for the vdW interaction between uniaxially strained graphene
and several types of atoms (with different masses and polarizibilities) as a function of 
strain and the electron interaction coupling constant.
Section~\ref{sec:qr} contains results for the elastic  quantum reflection (QR) coefficient
as a function of strain and Section~\ref{sec:helium_4_adsorption} discusses
the many-body adsorption potential for helium on strained graphene.
In Section~\ref{sec:conclusions} we present our conclusions and
perspectives for further exploration.

\section{Atom-Graphene van der Waals Force}
\label{sec:atom_graphene}
We begin with the problem of strain-depenendence of the atom-graphene van der Waals  potential.
The  theory of vdW forces is described in Refs.~[\onlinecite{ll1,ll2,vdw1}]
and contains, in particular, the fully relativistic treatment within 
Lifshitz theory, which amounts to the well-known Random Phase
Approximation (RPA), including retardation effects incorporated through the
polarization function and the interactions. Many works have also been devoted
to atom-graphene interactions  \cite{vdw3, AVDW-2, vdw5, vdw4, AVDW-5,vdw2}.
As is well known, and we will see explicitly, relativistic effects depend on
the interaction distance and are relatively weak on the nanometer scale (only becoming important
on micron scales).   Thus we find it useful to write down the (less cumbersome)
non-relativistic expressions first, and then include relativistic effects.
The zero temperature formalism is used since finite temperature effects
are not important in the small distance regime under consideration.

\subsection{Non-relativistic treatment}
 
The dynamic atomic polarizability $\alpha(i\omega)$ for various atoms, which
is required for the calculation, is known with great precision \cite{atomic},
and for most atoms can be approximated by the following single-oscillator form
(for the vdW force, one needs it on the imaginary axis): 
\begin{equation}
\alpha(i\omega) = \frac{\alpha_0 \omega_0^{2}}{\omega_0^{2}+\omega^{2}}.
\label{eq:alpha}
\end{equation} 
Here $\alpha_0$ is the static polarizability. We have performed
detailed fits of this form to the data of Ref.~[\onlinecite{atomic}] for three  atoms,
and our results are in very good agreement with parameter values quoted in the literature.
 We obtain, for  $ \text{H: } \alpha_0= 4.5  \text{ a.u.},  \ \omega_0=11.65  \text{ eV}$; for $\text{Na, }
\alpha_0= 162.6 \text{ a.u.}, \ \omega_0=2.15 \text{ eV}$, and for  $
\text{He, } \alpha_0= 1.38 \text{ a.u.},  \ \omega_0=27 \text{ eV}.$ These
atoms were  chosen because their behavior is relevant to cold atom
experiments. Notice that they have very different polarizabilities, where the  
atomic unit of polarizability is $1\text{ a.u.} = 1.4818 \times 10^{-4}
\text{ nm}^3$.
 
Next, the polarization of the graphene electrons is needed. Assuming uniaxially
strained graphene, as shown in Fig.~\ref{fig:strain}, for weak to moderate
strain the electronic dispersion is well described by an effective anisotropic
Dirac dispersion  $E({\bf k})$
with  different, strain-dependent velocities $v_x, v_y$
\cite{pereira, choi,vdw-anand}:
\begin{equation}
E({\bf k}) ^2 = v_{x}^2 k_x^2 + v_{y}^2 k_y^2.
\label{dispersion}
\end{equation}
As mentioned previously, for a lattice deformation in the armchair direction,
the system remains semi-metallic (no gap opens) even for strong strain
\cite{pereira, choi}. For strain in other directions, in particular in the
zig-zag direction, a gap eventually opens as a function of strain,  the Dirac
cones become severely distorted (merging at the transition point) and cannot be
described by Eq.~\eqref{dispersion}.  Returning to the case of armchair strain,
we have performed a fit to the data described in
Refs.~[\onlinecite{choi,pereira}], which gives the anisotropy ratio $v_y/v_x$
as a  function of strain $\delta$.  The relationship between $v_y/v_x$ and
$\delta$ will be needed in Section~\ref{sec:helium_4_adsorption}.  We assume
strain to be in the $y$-direction (armchair direction), reducing the corresponding velocity while
the velocity in the perpendicular ($x$) direction is not significantly affected
\cite{pereira, choi}.  As explained in those works,  for small strain $\delta$
the variation of the velocities is  linear, and we find that for the armchair
direction is described well by the formulas: $v_y/v_F = 1 - \Lambda \delta, \
\  v_x/v_F = 1 +\Lambda \nu \delta$. Here $\nu = 0.165$ is the Poisson ratio,
$\Lambda \approx 2.23$, and $v_F$ is the velocity of unstrained graphene.
Beyond weak ($\approx10 \%$) strain, the dependence on strain becomes (only weakly)
nonlinear, and the above formulas  continue to approximately describe 
the numerical results \cite{choi} even for moderately strong deformations.
After taking into account the weak non-linearity in the armchair direction, we
arrive at the correspondence between velocity anisotropy and strain shown in
Table~\ref{tab:velstrain}.
\begin{table}[h]
\begin{center}
    \renewcommand{\arraystretch}{1.6}
  \begin{tabular}{  c  c  c  c  c }
   \hline\hline 
    $v_y/v_x$ & $1.00$ & $0.75$ & $0.40$ & $0.20$ \\ 
    $\delta$ & $0.00$ & $0.10$ & $0.25$ & $0.34$ \\ 
    \hline \hline
  \end{tabular}
\end{center}
\caption{\label{tab:velstrain}The relationship between the Fermi velocity anisotropy and
    the elongation of the $y$-axis of a strained graphene lattice using data
inferred from Refs.~[\onlinecite{choi,pereira}].}
\end{table}

For the rest of this section we will vary the effective Dirac
anisotropy $v_y/v_x$ from 1 (isotropic graphene) down to its largest value of
0.2 and we introduce the notation:
 \begin{equation}
 v_{\perp} \equiv \frac{v_y}{v_x} \leq 1.
 \end{equation}
 We consider graphene at half-filling, i.e. the chemical potential is at zero
 (the lower Dirac cone is full, the upper one is empty).  Returning to the
 calculation of the polarization, a simple rescaling of the isotropic graphene case leads to the exact expression \cite{vdw-anand}:
\begin{equation}{\label{e2}}
\Pi(\textbf{q},i\omega) = -\frac{1}{4 v_{x} v_{y}}\frac{v_{x}^2 q_x^2 + v_{y}^2 q_y^2}{\sqrt{v_{x}^2 q_x^2 + v_{y}^2 q_y^2 +\omega^2}}.
\end{equation} 

\noindent
From here, the vdW energy  in the non-relativistic limit is \cite{vdw1}:
\begin{eqnarray}
U_{\text{vdW}}(z) & = & - \frac{\hbar}{2\pi}\int_{0}^{\infty} d\xi \alpha(i\xi) \    2 \int_{0}^{\infty}dk  k^2 e^{-2kz} \times \nonumber \\
&&\int_{0}^{2\pi}
\frac{d\phi}{2\pi} \  \frac{|V(\textbf{k})\Pi(\textbf{k},i\xi)|}{1- V(\textbf{k})\Pi(\textbf{k},i\xi)}.
\label{eq:Vz1}
\end{eqnarray}
This is the RPA result.
Here $\phi$ is the angle between the $k_x$ and $k_y$ directions (in the strained case there is an explicit angular dependence).
The Coulomb potential is: 
\begin{equation}
V(\textbf{k}) = \frac{2\pi e^2}{k},  \   \  k=|\textbf{k}|.
\end{equation}
Finally, we define graphene's dimensionless coupling constant $g$ as:
\begin{equation}
g = \frac{\pi}{2}\frac{e^2}{v_x} 
\end{equation}
where we have set $\hbar=1$.  The value in vacuum can be obtained by
noting that for graphene ${e^2}/{v_x} \approx 2.2$, \cite{kotov} leading to
$g\approx 3.45$. If graphene is placed on a substrate (and one assumes 
vacuum in the upper half-space), the effective charge $e^2$ decreases due to
the dielectric constant, $\kappa$, of the substrate and we have to replace $e^2
\rightarrow 2 e^2/(1+\kappa)$.  For example, a SiO$_2$ substrate 
has $\kappa\approx 4$ and the coupling $g$ decreases substantially
\cite{kotov}.

Returning to Eq.~\eqref{eq:Vz1}, the vdW potential can be conveniently expressed  as:
\begin{equation}
    U_{\text{vdW}}(z) = - \frac{C_{3}(z)}{z^3},
\label{eq:VzC3}
\end{equation}
where:
\begin{eqnarray}
C_{3}(z) &=& \frac{\alpha_0 \omega_0}{8\pi}\frac{g}{v_{\perp}} \int_{0}^{\infty} \frac{d\omega}{1+\omega^2}
\int_{0}^{2\pi}\frac{d\phi}{2\pi}  
\int_{0}^{\infty}dq q^3 e^{-q} \times \nonumber \\
&&\frac{f(\phi,v_{\perp})}{\sqrt{q^2f(\phi,v_{\perp}) + \omega^2\Omega^2} +
(g/v_{\perp})qf(\phi,v_{\perp}) }.
\label{C3}
\end{eqnarray}
Here we  have written the result in such a way that the physical dimension of
$C_{3}$ comes only from the pre-factor  $\alpha_0 \omega_0$ , while the other
couplings and the integration variables $q,\omega$ are dimensionless.
We also use the definition:
\begin{equation}
f(\phi,v_{\perp}) = \cos^2{\phi} + v_{\perp}^2 \sin^2{\phi}.
\end{equation}
The characteristic dimensionless scale $\Omega$ is defined as 
\begin{equation}
\Omega = \Omega(z) \equiv \frac{2\omega_{0} z}{v_x},
\end{equation}
and is distance-dependent.

\subsection{Relativistic effects}
\label{ssec:relativistic}

The above formulas are generalized to take into account relativistic
corrections in full, which enter in two ways.  First, there is
an explicit contribution from retarded potential pieces
\cite{vdw1,vdw2,vdw3,vdw4,ll1,ll2}, proportional to $(v_x/c)^2 = (1/300)^2 \ll
1$, which can be safely neglected. Second, there is a  retardation
modification of  the Coulomb interaction portion, which  now reads
(as before, all integration variables are dimensionless)
\begin{eqnarray}
C_{3}(z)&=& \frac{\alpha_0 \omega_0}{8\pi}\frac{g}{2v_{\perp}} \int_{0}^{\infty} \frac{d\omega}{1+\omega^2}
\int_{0}^{2\pi}\frac{d\phi}{2\pi} \int_{\omega_c \omega}^{\infty}dq q^3 e^{-q} \times \nonumber \\
&&\frac{f(\phi,v_{\perp}) \times \left ( 2 - \frac{\omega_c^2\omega^2}{q^2}\right )}{\sqrt{q^2f(\phi,v_{\perp})+ \omega^2\Omega^2} + (g/v_{\perp})qf(\phi,v_{\perp}) },
\label{C3rel}
\end{eqnarray}
where we introduce the relativistically generated dimensionless,
distance-dependent  scale, which in particular  provides an effective cutoff in
the above integration:
\begin{equation}
\omega_c \equiv\Omega(z)/(c/v_x) = \Omega(z)/300.
\end{equation}
The non-relativistic formula is recovered for $\omega_c=0$ ($c=\infty$).
It  is clear that for  finite speed of light $c$ the relativistic effects become more important as the distance, $z$, increases
(so that $\omega_c $ starts deviating substantially  from $0$).  At very  large $z$, within the regime $\omega_c 
>  \Omega$, $C_3(z) \sim 1/z$, i.e. the vdW potential changes shape. Our approach above is equivalent  (apart from different notation)
to the conventionally used Lifshitz theory \cite{ll1,ll2,vdw1}, and the $C_3$ results for
isotropic graphene are completely consistent with  published  numbers for H, He, Na 
 \cite{vdw3, AVDW-2, vdw5, vdw4, AVDW-5,vdw2}.

% Results

\begin{figure}[t]
\begin{center}
\includegraphics[width=1.0\columnwidth]{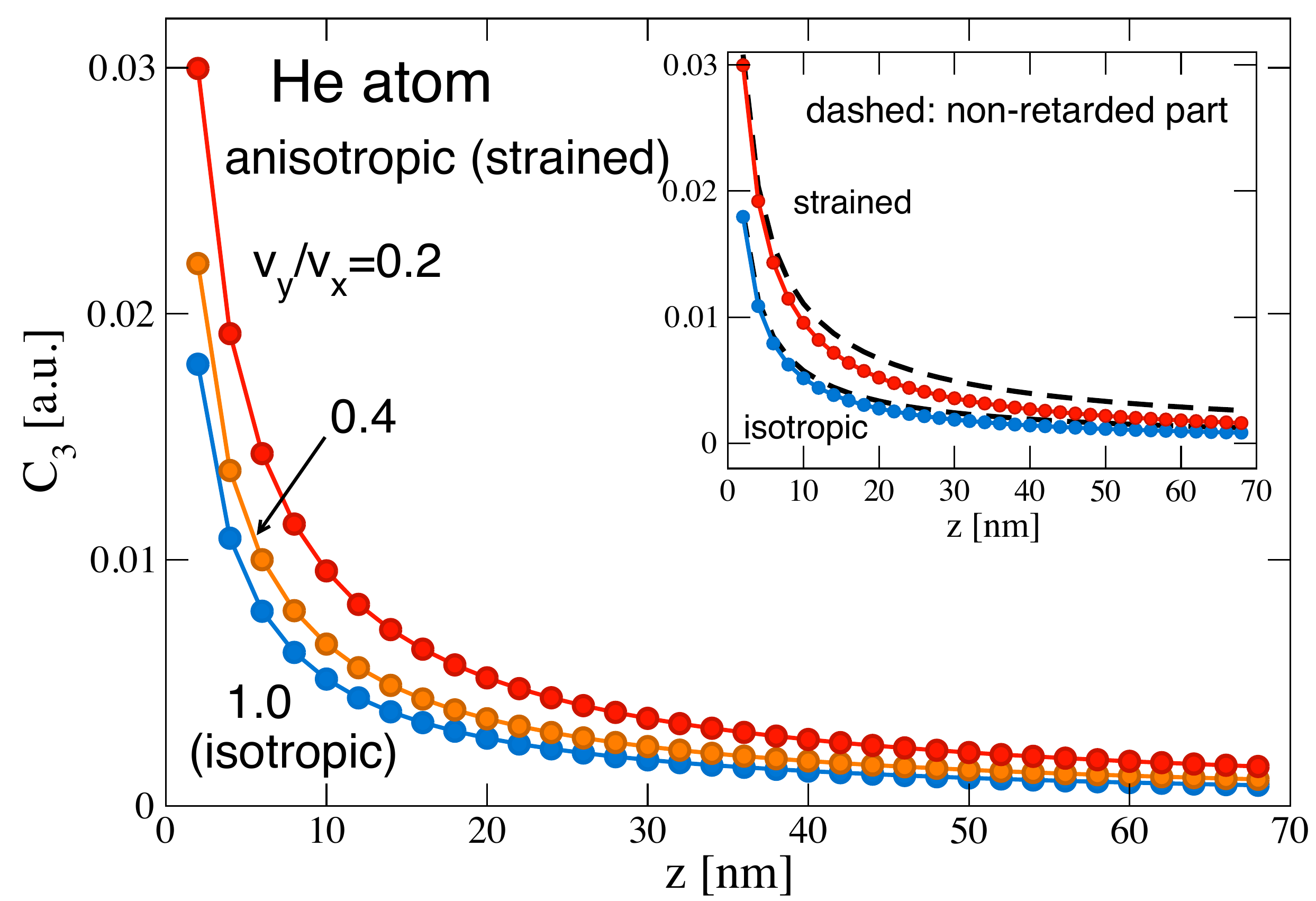}
\end{center}
\caption{(Color online)
Distance dependence of $C_3(z)$ which determines the vdW potential via
Eq.~\eqref{eq:VzC3}.  Results are plotted for a He atom, for several values of
the anisotropy $v_{\perp}=v_y/v_x$.  Inset: Comparison of the fully
relativistic expression  with the non-relativistic formula (dashed lines) for
the isotropic ($v_{\perp}=1$) and maximally strained ($v_{\perp}=0.2$) cases.}
\label{fig:helium}
\end{figure}

Now, we proceed to a more detailed discussion of our results as a function of
strain (Dirac cone anisotropy).  First we show that $C_3(z)$  generically
has substantial distance dependence, which is already present in the
non-relativistic  limit, Eq.~\eqref{C3}, due to the  frequency dependence
(semi-metallic nature) of graphene's polarization.  A comparison of
Eq.~\eqref{C3} with the fully relativistic expression Eq.~\eqref{C3rel} is
presented in Fig.~\ref{fig:helium}~(inset), for He. The difference between the
two is appreciable (even in the nm distance range), and increases with
distance, as expected.

Therefore, in order to achieve maximum accuracy, from
now on we will use the fully relativistic formula Eq.~\eqref{C3rel}.  The main
panel in Fig.~\ref{fig:helium} shows the dependence of $C_3$ on the anisotropy
$v_{\perp}$ for He. We assume, for definitiveness,  that graphene is
free-standing (in vacuum)  i.e. the electron-electron coupling is $g=3.45$. 
%
% ------------------------------------------------------------------------------- 
% ------------------------------------------------------------------------------- 
%
Atomic units of $C_3$ are defined as: $1\text{ a.u.  of } C_3 =
\SI{4.032}{\milli\electronvolt.\nano\meter^3}$.  We find a significant
dependence on strain, which tends to increase the value of $C_3$. This increase
can be traced to the enhancement of the electron polarization from Eq.~\eqref{e2}
with strain.  A factor of $2$ increase in the vdW potential is seen for almost all distances at the maximal strain under consideration ($v_{\perp}=0.2$).
Finally, the shape of the curves in  Fig.~\ref{fig:helium}  suggests that the
vdW potential experiences significant deviations from  a pure $1/z^3$ tail, even
at such intermediate (nm) distances. We will analyze this crossover at the end
of this Section.

\begin{figure}[t] \begin{center}
\includegraphics[width=1.0\columnwidth]{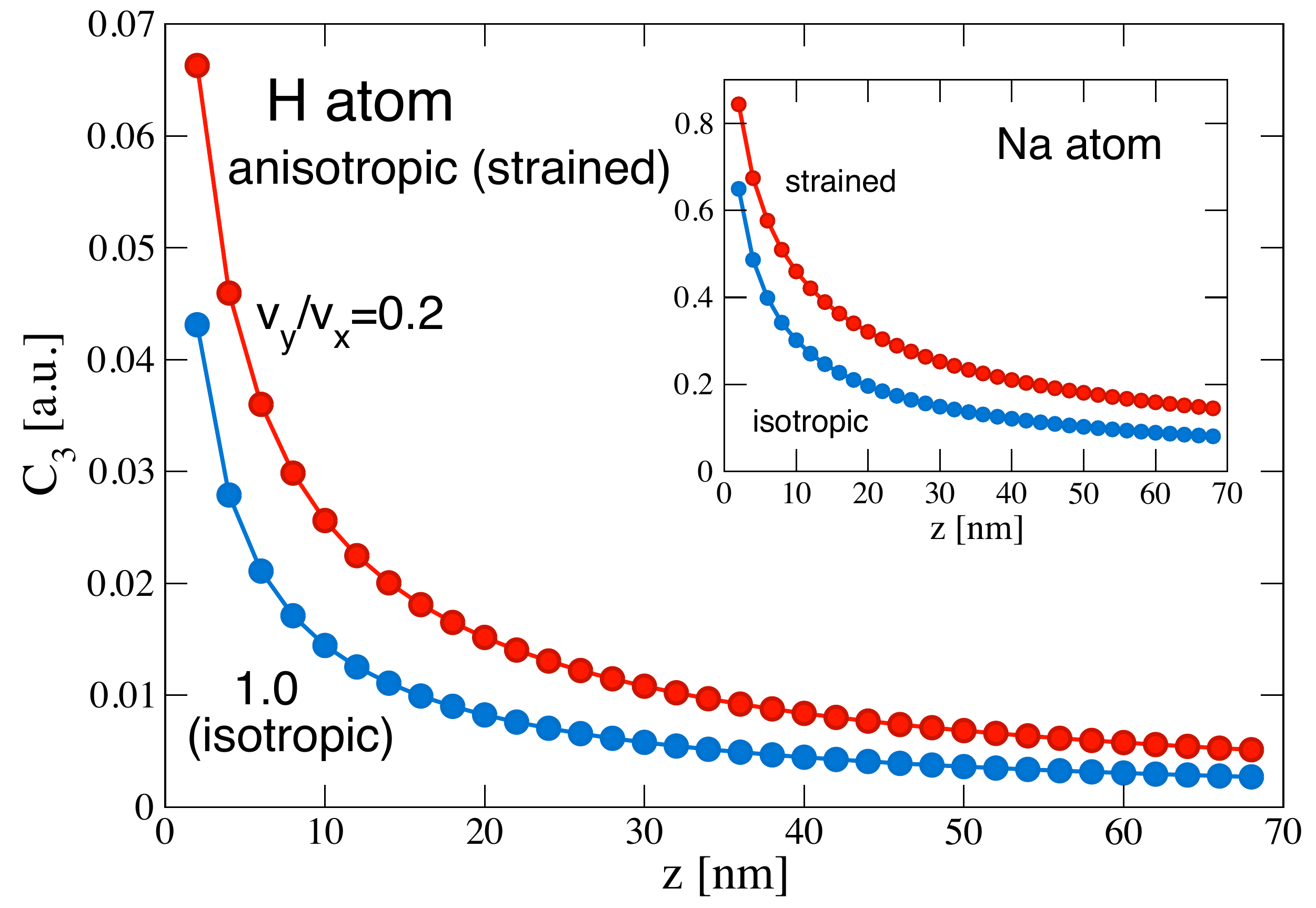} \end{center}
\caption{(Color online) Distance dependence of $C_3(z)$ which determines the
    vdW potential via Eq.~\eqref{eq:VzC3}, for H atom on isotropic
    ($v_{\perp}=1$) and maximally strained graphene ($v_{\perp}=0.2$).  Inset:
    Same as main panel, for a Na atom; note the larger vertical scale.} 
\label{fig:h-na} 
\end{figure}

In Fig.~\ref{fig:h-na}, we present our results for H and Na atoms. 
%
% ------------------------------------------------------------------------------- 
% ------------------------------------------------------------------------------- 
%
Even though the scales of $C_3$ differ significantly, due to the very different
atomic polarizabilities (Na is approximately forty times more polarizable), the overall
anisotropic behavior for these atoms is quite similar. It is also similar to the
case of (weakly polarizable) He shown in  Fig.~\ref{fig:helium}.

\begin{figure}[t]
\begin{center}
\includegraphics[width=1.0\columnwidth]{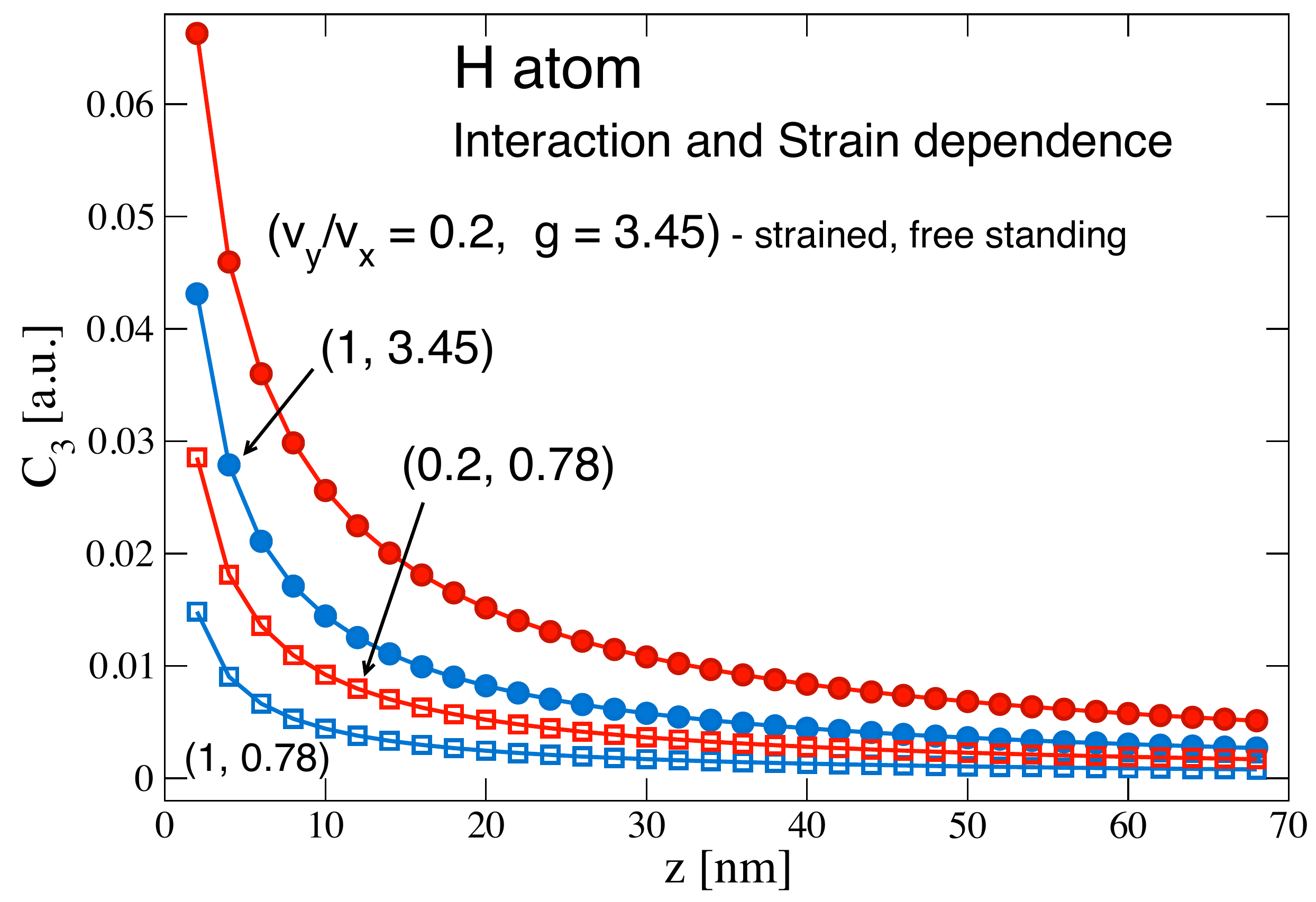}
\end{center}
\caption{(Color online)
Combined correlation ($g=(\pi/2)(e^2/v_x)$) and anisotropy ($v_{\perp}$)  dependence of
$C_3(z)$ for a H atom. The  representative electron interaction values correspond to $e^2/v_x=2.2$  (vacuum),
and $e^2/v_x=0.5$ (screened case).}
\label{fig:h-coulomb}
\end{figure}

In Fig.~\ref{fig:h-coulomb}, we plot results for the combined effect of
anisotropy and electron-electron interaction $g$.  
%
% ------------------------------------------------------------------------------- 
% ------------------------------------------------------------------------------- 
%
The interaction controls
both the  overall scale of $C_3$ and  (metallic) screening, as reflected in the
denominator of Eq.~\eqref{C3rel}. If screening were absent, the strain
dependence of $C_3$ would be much more pronounced.  This effect is only
marginally visible in Fig.~\ref{fig:h-coulomb}, i.e. the increase of $C_3$ is
slightly larger for $g=0.78$ (lower two curves) than for $g=3.45$ (upper two
curves).  The overall reduction of the vdW interaction as $g$ decreases is the
dominant behavior.

\begin{figure}[t]
\begin{center}
\includegraphics[width=1.0\columnwidth]{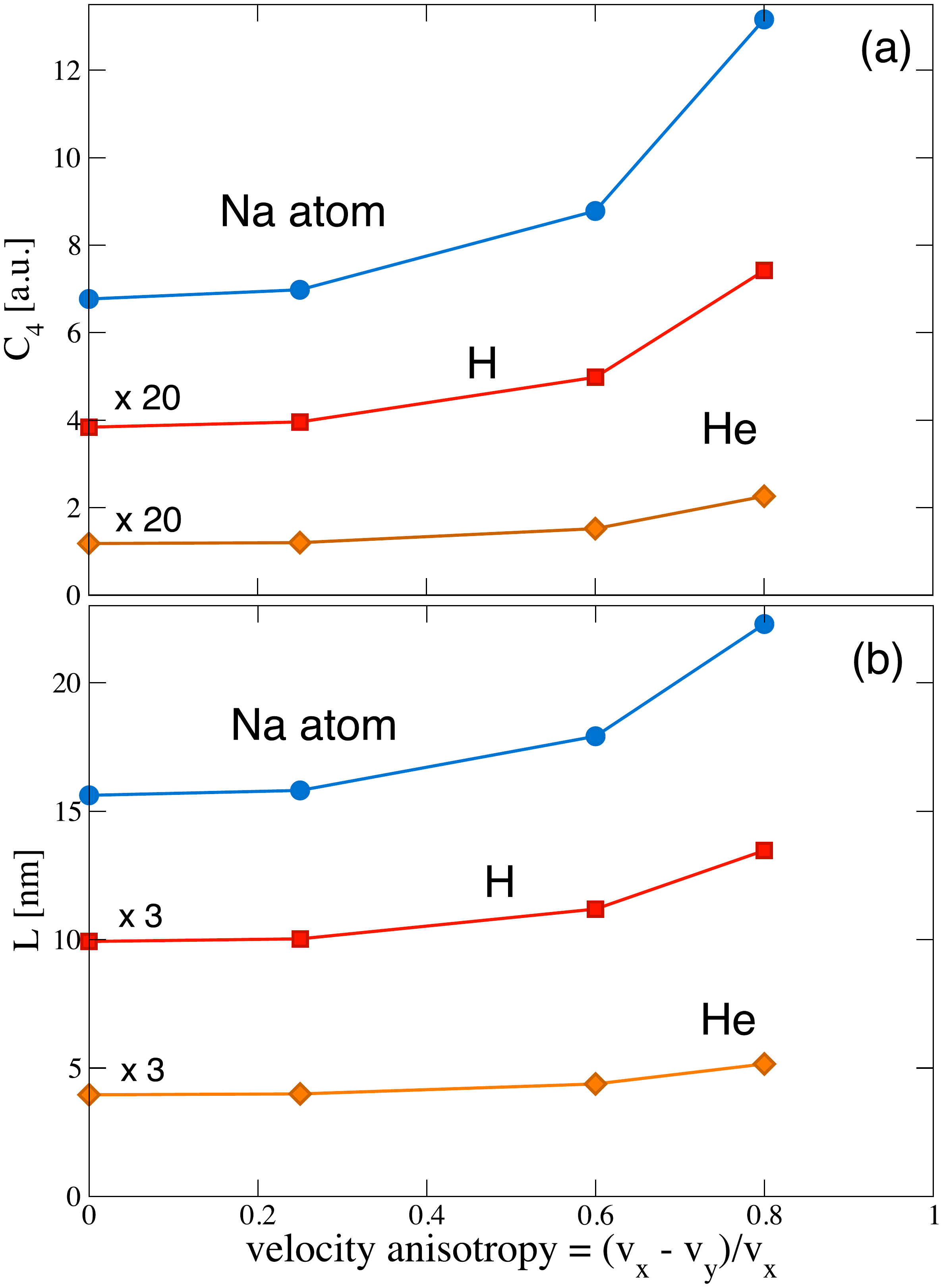}
\end{center}
\caption{(Color online)
Crossover from $C_3/z^3$ to $C_4/z^4$ behavior in the vdW potential tail,
as a function of the anisotropy $(1-v_{\perp})$. 
Fits are performed to the expression $C_3(z) = C_{4}/(z+L)$ with panel (a) showing
$C_4$ and (b) the value of the crossover lengthscale $L$.}
\label{fig:c4}
\end{figure}

Finally, we examine the crossover in the distance dependence of the potential.
The significant  $z$-dependence of $C_3(z)$ suggests a fit of the form
$C_3(z) = C_{4}/(z+L)$, where $C_4$ and $L$ are the fitting parameters.  In
standard atomic units,  $1\text{ a.u.  of } C_4 =
\SI{4.032}{\milli\electronvolt.\nano\meter^4}$.  Our results for all studied
atoms are summarized in Fig.~\ref{fig:c4}. 
%
% ------------------------------------------------------------------------------- 
% ------------------------------------------------------------------------------- 
%
We find that the crossover distances are in the nm distance
region, and increase with strain. For He and H  we have $L \simeq 1.3,
\SI{3.3}{\nano\meter}$
respectively, for isotropic graphene, while the corresponding value for Na is
significantly larger, $L \simeq \SI{15.6}{\nano\meter}$.  Thus at distances $z \gg L$ the vdW
potential becomes $U_{\text{vdW}}(z) = - C_{4}/z^4$.  The coefficient $C_4$
also increases with strain as shown in  Fig.~\ref{fig:c4}.

\section{Implications for Quantum Reflection}
\label{sec:qr}

As a first application of our results, we consider the strain dependence of the
quantum reflection (QR).  For elastic interactions, ultracold atoms impinging on graphene should be subject to quantum reflection
from the attractive vdW tail of the atom-graphene potential.  QR is a simple
result of the wave-like nature of low-energy particles moving in an attractive
potential that falls off sufficiently rapidly with distance from the surface.
Under QR, an ultracold atom can have a high probability of reflecting without
ever reaching a classical turning point near the graphene surface.   Studied since the
development of quantum mechanics, QR continues to fascinate both theorists
\cite{qr-th, anti-h, Dennis} and experimentalists \cite{pasquini04, pasquini06,
qr-exp} alike and while QR has been previously studied for graphene \cite{qr-th}, we
now investigate the effect of uniaxial graphene strain on its properties. Since we have
found  in the previous Section that the vdW potential is sensitive to the Dirac
anisotropy,  this implies that QR might be efficiently tuned with uniaxial
strain.

To determine how the vdW potential affects above-barrier quantum reflection
(see e.g.  Refs.~[\onlinecite{friedrich1,friedrich2}] for an overview) we
consider a non-relativistic atom with energy $E=\hbar^2k^2/(2M)$ and mass $M$
impinging on graphene (where we have temporarily restored $\hbar$ for clarity).
The behavior of the QR reflection coefficient $R$, defined as the magnitude of
the reflectivity (i.e. the piece of the
wave-function which is reflected), depends on the distance  dependence of
$U_{\text{vdW}}(z)$. In the regime under consideration where
$U_{\text{vdW}}(z)$ experiences a crossover from  $-C_3/z^3$ to  $-C_4/z^4$
behavior, the  value of the effective parameter $\rho$  determines which part of the
tail is the dominant contribution to $R$:\cite{friedrich1}
\begin{equation}
\rho =\frac{\sqrt{2M}}{\hbar}\frac{C_3}{\sqrt{C_4}},
\end{equation}
where $C_4$ is determined from Fig.~\ref{fig:c4} and the constant $C_3$ is
defined as $C_3\equiv C_4/L$.  In the low-energy regime, if  $\rho \ll 1$, then  the
$-C_3/z^3$ part determines $R$, while $\rho \gg 1$ means that the $-C_4/z^4$
tail is more important.  Taking into account our results from Fig.~\ref{fig:c4},
we find  the following numbers for different atoms (for isotropic graphene):
$\rho_{\text{H}} \approx 1.9$, $\rho_{\text{He}} \approx 5.2$ and
$\rho_{\text{Na}} \approx 11.2$.  The value for H is the smallest, resulting from its small atomic mass.  From the results of Ref.~[\onlinecite{friedrich1}],
we can see that the value of $\rho$ that separates the asymptotically small and
large values is around $\rho \approx 3$. 

Let us consider, for definitiveness, the case of Na, where
the $-C_4/z^4$  tail is dominant, and estimate the effect of strain on $R$.
It is convenient to define  \cite{friedrich1} the length scale $\beta_4$ via:
\begin{equation}
    U_{\text{vdW}}(z) = -\frac{C_4}{z^4} \equiv -
    \frac{\hbar^2}{2M}\frac{\beta_4^2}{z^4}, 
\end{equation}
such that
\begin{equation}
    \beta_4=\frac{\sqrt{2MC_4}}{\hbar}.
\end{equation}
The asymptotic behavior of $R$ in the low-energy region $E \rightarrow 0$,
or in proper dimensionless units, $\beta_4 k \ll 1$, is then:\cite{friedrich1}
\begin{equation}
R \approx 1 - 2 (\beta_4 k) ,  \ \ \beta_4 k \ll 1.
\end{equation}
In the opposite, high-energy regime, we have:
\begin{equation}
R \sim  e^{-1.694  \sqrt{\beta_4 k}}, \  \  \  \beta_4 k \gg 1,
\end{equation}
which is valid, provided $1 \ll \beta_4 k \ll \rho^2$. \cite{friedrich1}

\begin{figure}[t]
\begin{center}
\includegraphics[width=1.0\columnwidth]{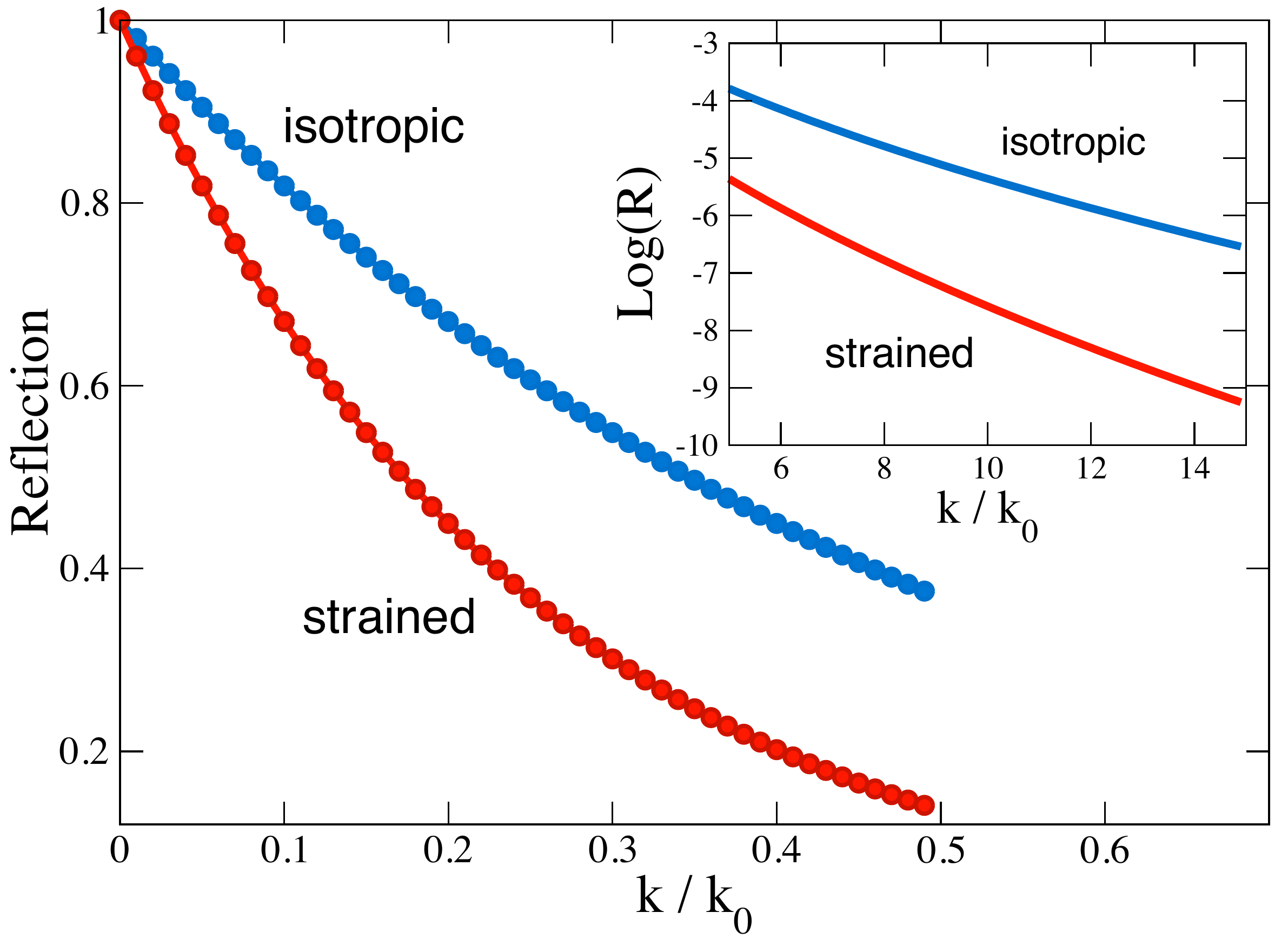}
\end{center}
\caption{(Color online)
Asymptotic behavior of the quantum reflection coefficient $R$ for unstrained graphene
and anisotropic (strained) graphene with $v_{\perp}=0.2$, in the regime
where the tail $-C_4/z^4$  dominates, as a function of the atomic
momentum $k$. The length scale is defined as $1/k_0 \equiv \beta_4$.
In the main panel the low energy behavior is plotted. Inset: High energy
behavior, corresponding to exponentially small reflection.}
\label{fig:qr}
\end{figure}

Since $C_4$ (and therefore $\beta_4 $) increases with
strain, (see Fig.~\ref{fig:c4}), it is clear that larger strain leads to a decrease of the quantum
reflection, as shown in Fig.~\ref{fig:qr}. This decrease of the QR can be very
substantial (for moderately large strain). Finally, in
situations where the $-C_3/z^3$ piece of the vdW tail dominates, the
corresponding asymptotic behavior is also well established \cite{friedrich1}
and the strain dependence can be readily calculated, leading to behavior
qualitatively very similar to the one in  Fig.~\ref{fig:qr}.
%
% ------------------------------------------------------------------------------- 
% ------------------------------------------------------------------------------- 
%

Having understood that uniaxial strain enhances the van der
Waals interaction between an impinging atom and a deformed
graphene substrate, thus leading to a marked reduction in the probability of
its reflection, we now ask what effects it may have
\emph{near} the surface.  In particular, we investigate the physics of
adsorption of light atoms onto mechanically strained graphene.

\section{Helium-4 adsorption potential}
%======================================
\label{sec:helium_4_adsorption}

In this section we focus exclusively on the interaction between a single
$^4$He atom and graphene, but the techniques we develop could be
applied to the study of adsorption of any neutral polarizable atom.

One conventional treatment of helium adsorption on graphene (or a graphite
surface) \cite{Carlos:1980cf} estimates the
total potential energy $U(\vec{r})$ for a neutral adatom at position $\vec{r} =
(x,y,z)$ as a discrete summation of Lennard-Jones (6--12) two-body interactions with
the $N$ carbon atoms located at $\vec{R}_i = (X_i,Y_i,0)$:
\begin{equation}
    U(\vec{r}) = 
    4 \varepsilon \sum_{i=1}^N \left[
        \left|\frac{\sigma}{\vec{r}-\vec{R_i}}\right|^{12} - 
    \left|\frac{\sigma}{\vec{r}-\vec{R_i}}\right|^{6} \right].
\label{eq:Usum}
\end{equation}
We note that the adatom experiences the effects of a corrugated graphene sheet
at short distances and thus the potential is a function of the full spatial
coordinate $\vec{r}$ as opposed to the continuum approximation used
in Eq.~\eqref{eq:VzC3} where it is only sensitive
to the height $z$ above the sheet.  The $r^{-12}$ form of the short distance
interaction is semi-empirical and is meant to capture the effects of Pauli
repulsion from overlapping electronic orbitals,  while the $r^{-6}$ attractive
part of Eq.~\eqref{eq:Usum} is due to the individual vdW dispersion forces
between the neutral carbon and helium atoms. For two interacting atoms, the
Lennard-Jones (LJ) parameters $\sigma$ and $\varepsilon$ set the location of
the minimum at $r_m = 2^{1/6} \sigma$ and its depth at $-\varepsilon$.  For
pure gases and liquids, they can be estimated using second-virial or viscosity
coefficients \cite{Oobatake:1972je} whereas for mixtures, they can be roughly
approximated \cite{Boda:2008yy} using the Lorenz-Bertholot mixing rules, which
for two species $A$ and $B$ are given by:
\begin{equation}
\begin{aligned}
\varepsilon_{A-B} &= \sqrt{\varepsilon_A \varepsilon_B} \\
    \sigma_{A-B} &= \frac{\sigma_A + \sigma_B}{2}.
\end{aligned}
\end{equation}
For a single helium atom interacting with carbon in either graphene or
graphite, the most commonly used parameters are taken from
Ref.~[\onlinecite{Carlos:1980cf}] to be $\varepsilon_{\mathrm{He}-C} =
16.2463$~K and $\sigma_{\mathrm{He}-C}=
2.74$~\AA. These values were determined by comparing the bound states of
Eq.~\eqref{eq:Usum} to experimental results for the adsorption spectra of helium
on graphite \cite{Derry:1979sa}.   It is thus natural to ask if these LJ
parameters can be used to capture the effects of strain considered in the
previous sections within the continuum approximation and shown in
Fig.~\ref{fig:helium}. The answer to
this question constitutes the remainder of this paper.

\subsection{Unstrained graphene}
%--------------------------------------------
\label{sub:unstrained_graphene}

\begin{figure}[t]
\begin{center}
\includegraphics[width=1.0\columnwidth]{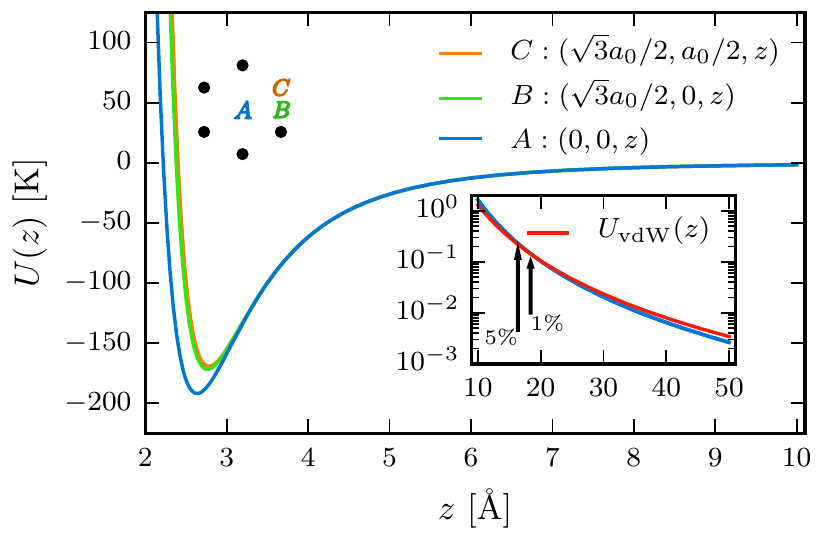}
\end{center}
\caption{(Color online) The Lennard-Jones potential for a single helium atom
    located a distance $z$ above a graphene sheet at positions $A,B,C$ 
    (shown in upper left) using $\varepsilon_{\mathrm{He}-C} = 16.2463$~K and
    $\sigma_{\mathrm{He}-C}= 2.74$~\AA. The inset (same axes) shows a
    comparison with the long distance van der Waals potential computed 
    using the continuum polarization of the graphene sheet.  Arrows indicate the
values of $z$ where the relative error  between the two calculations is 5\% and
1\% respectively.} \label{fig:VzIsoNoFit}
\end{figure}

We begin our analysis by investigating the accuracy of the Lennard-Jones
potential for helium interacting with an isotropic graphene sheet by comparing
Eq.~\eqref{eq:Usum} with the long-distance continuum limit value
$U_{\text{vdW}}(z) = -C_3(z)/z^3$ in Eq.~\eqref{eq:VzC3}. In
Fig.~\ref{fig:VzIsoNoFit} we show the adsorption potential for $N=2^{18}$ carbon
atoms using the standard LJ parameters for He-C interactions as a function of
distance above the graphene sheet for the three high symmetry locations
$\vec{r}_A = (0,0,z)$, $\vec{r}_B = (\sqrt{3}a_0/2,0,z)$ and
$\vec{r}_c=(\sqrt{3}a_0/2,a_2/2,z)$, shown in the upper left inset. Here  
$a_0=\SI{1.42}{\angstrom}$ is the isotropic C-C bond length.  
%
% ------------------------------------------------------------------------------- 
% ------------------------------------------------------------------------------- 
%
The main panel depicts the usual form of the LJ adsorption potential at short
distances, with the details of the attractive minima and hardcore repulsion
depending on the relative orientation of the adatom with respect to the
graphene lattice \cite{Carlos:1979vj}.  For distances
$z>\SI{10}{\angstrom}$, the potential is insensitive (at the order of
\SI{E-10}{\kelvin}) to the $x$ and $y$ positions of the adatom and the
substrate can be effectively treated in the continuum approximation. 
The inset shows $U(\vec{r}_A)$ along with the
continuum long distance calculation $U_{\rm vdW}(z)$ for He from
Section~\ref{ssec:relativistic} with the relative error
decreasing from $5\%$ at $z\simeq\SI{16}{\angstrom}$ to $1\%$ at
$z\simeq\SI{18}{\angstrom}$.  We stress that
this agreement is achieved with no adjustable parameters and serves as an
excellent benchmark of our continuum calculations at long distances.  

\subsection{Lennard-Jones parameters for strained graphene}
%---------------------------------------------------------------------------
\label{sub:lennard_jones_parameters_for_strainedgraphene}

In Section~\ref{sec:atom_graphene} we found that the dispersion force between
adatoms and graphene \emph{increases} at long distances as a function of
increasing mechanical strain.  This finding can be investigated by evaluating
the discrete LJ potential for graphene lattices with strain $\delta =
0.0,0.1,0.25,0.34$ defined as the relative elongation of the lattice along the armchair
direction and corresponding to the velocity anisotropies $v_y/v_x$ considered
above and shown in Table~\ref{tab:velstrain}.  For each value of the strain
parameter $\delta$ we construct a graphene lattice consisting of $N = 2^{18}$
atoms in the $z=0$ plane with positions defined by the lattice: 
\begin{equation}
\begin{aligned}
    \vec{a}_1 &=\frac{\sqrt{3}a_0}{8}(4+\delta-3\delta\nu,\sqrt{3}(4+3\delta-\delta\nu)) \\
    \vec{a}_2
    &=\frac{\sqrt{3}a_0}{8}(-4-\delta+3\delta\nu,\sqrt{3}(4+3\delta-\delta\nu))
\end{aligned}
\label{eq:latticeVectors}
\end{equation}
and basis vectors:
\begin{equation}
\begin{aligned}
    b_1&=(0,0) \\
    b_2&=a_0(0,1+\delta)
\end{aligned}
\label{eq:basisVectors}
\end{equation}
where $\nu=0.165$ is the Poisson ratio for graphite \cite{pereira}.

\begin{figure}[t]
\begin{center}
\includegraphics[width=1.0\columnwidth]{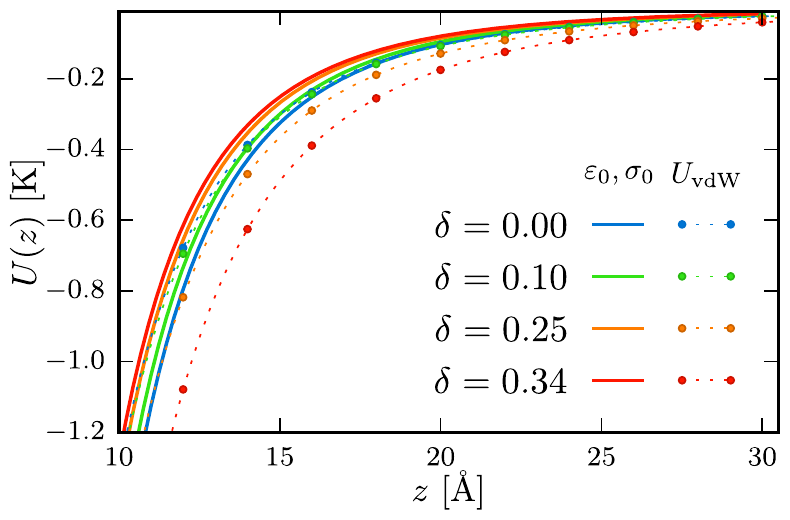}
\end{center}
\caption{(Color online) The long distance tail of the adsorption potential for
    a He adatom as a function of the height above a graphene sheet subject
    to mechanical strain parametrized by $\delta$.  Solid lines show the
    results of the discrete Lennard-Jones potential $U(z)$ using the standard
    parameters $\varepsilon_0,\sigma_0$ given in the text, while points
    connected by dashed lines show $U_{\rm vdW}(z)$ computed within a continuum
    approximation.  Note that the two methods show the opposite strain
    dependence indicating the failure of the standard Lennard-Jones calculation
for strained graphene.} 
\label{fig:nofit}
\end{figure}

For distances beyond $\SI{1}{\nano\meter}$, a comparison of the
discrete and continuum calculations for the potential for different strains
$\delta$ is shown in Fig.~\ref{fig:nofit} where we have again used the standard
LJ parameters for He--C, now labeling them $\varepsilon_0 \equiv
\varepsilon_{\mathrm{He}-C}$ and $\sigma_0 \equiv \sigma_{\mathrm{He}-C}$.
%
% ------------------------------------------------------------------------------- 
% ------------------------------------------------------------------------------- 
%
Not only do we find considerable disagreement away from $\delta=0$, but the
strain dependence has opposite signs; the discrete calculation yields
weaker dispersion forces as the strain is increased.  This finding indicates 
that the isotropic LJ parameters for He--C interactions cannot
be used when computing the adsorption potential for strained graphene lattices.
This failure is perhaps unsurprising, as these effective parameters are meant
to capture a plethora of microscopic details that are certainly strain dependent.

To address this fundamental discrepancy we have devised a procedure that
allows us to determine the strain dependence of $\varepsilon$ and $\sigma$. We
proceed by constructing a set of strained finite size graphene lattices, then 
compute the potential energy $U(z) \equiv U(0,0,z)$ using the brute-force
discrete summation in Eq.~\eqref{eq:Usum} 
for a fine mesh of LJ parameters $\varepsilon \in \{0.9\varepsilon_0, \cdots,
1.1\varepsilon_0\}$ and $\sigma \in \{0.9\sigma_0, \cdots, 1.1 \sigma_0\}$ with 
the expectation that the values of strain under consideration should not have a
an $\mathrm{O}(1)$ effect.  The resulting four dimensional data
set: $U\left(z;\varepsilon,\sigma,\delta\right)$ can then be compared with 
the long distance continuum value of $U_{\rm vdW}(z;\delta)$ using 
the mean squared residual:
\begin{equation}
    \chi^2(\varepsilon,\sigma,\delta) = \sum_{i=1}^{n} \left \lvert
    U(z_i;\varepsilon,\sigma,\delta) -
    U_{\text{vdW}}(z_i;\delta) \right \rvert^2
\label{eq:chi2}
\end{equation}
for $n=3364$ values of $z$ in the range $z =
\SIrange{16.37}{50}{\angstrom}$. The starting point for the residual calculation
of $z_0 = \SI{16.37}{\angstrom}$ was chosen such that the relative error between
the long distance potential and the discrete summation was equal to 5\% for
$\delta = 0$ (Fig.~\ref{fig:VzIsoNoFit} inset). Although this choice is
somewhat arbitrary, we found little dependence on the final results when
choosing $z_0 = \SI{18.40}{\angstrom}$ corresponding to a 1\% relative
error for isotropic graphene. We were, however, limited to single precision as
all computations were performed on graphical processing units to reduce their
run time. No finite-size effects were observed at this level for graphene
lattices with $N\ge2^{18}$ carbon atoms.  

\begin{figure}[t]
\begin{center}
\includegraphics[width=1.0\columnwidth]{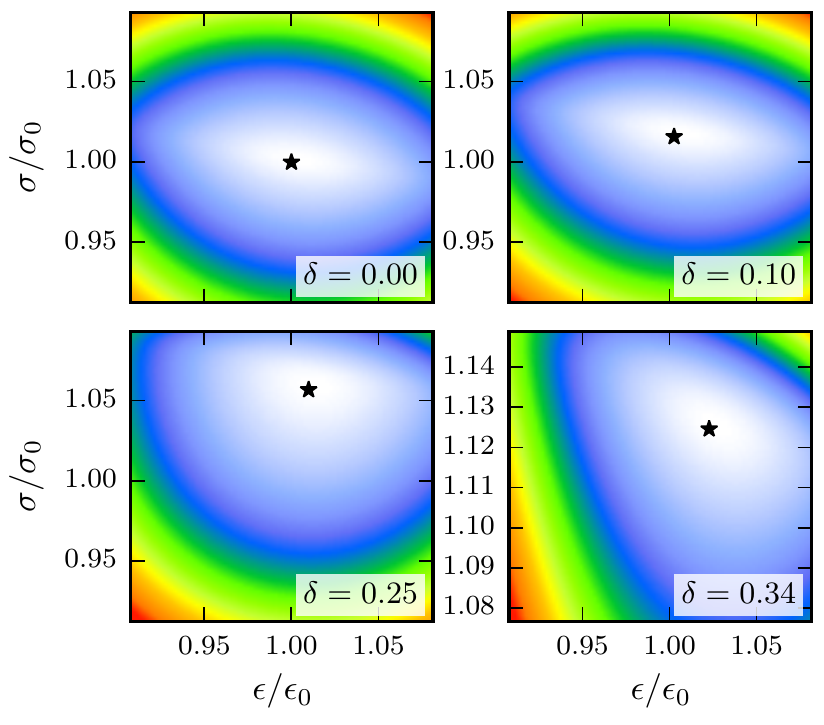}
\end{center}
\caption{(Color online) The likelihood estimator $\mathcal{S} \sim \chi^2 + \Delta^2$
    defined in Eq.~\eqref{eq:S} used to determine the value
of the Lennard-Jones parameters $\varepsilon$ and $\sigma$ producing the best
fit to the long distance continuum van der Waals potential for a single helium-4 atom
above strained graphene with $\delta = 0, 0.1, 0.25$ and $0.34$.  Axes are
normalized to the conventionally employed interaction parameters for helium and carbon: $\varepsilon_0 =
\SI{16.2463}{\kelvin}$ and $\sigma_0 = \SI{2.74}{\angstrom}$ with the density
scale indicating the goodness of fit from light (best) to dark (worst).  The
star indicates the identified global best fit.} 
\label{fig:chi2}
\end{figure}

The residual $\chi^2$ is minimized by the two-parameter function $\varepsilon(\sigma,\delta)$
over the range of parameters considered and we must add an additional constraint in order to extract the
optimal values of $\varepsilon$ and $\sigma$ for a given strain.  This can be accomplished by
requiring that the $\delta$-dependent LJ parameters are \emph{close} to the isotropic
ones $\varepsilon_0$ and $\sigma_0$.  To this end, we define a relative
Euclidean distance-cost function:
\begin{equation}
    \Delta^2(\varepsilon,\sigma,\delta) =
    \left(\frac{\varepsilon}{\varepsilon_0}-1 \right)^2 +
    \left(\frac{\sigma}{\sigma_0}-1 \right)^2
\label{eq:Delta}
\end{equation}
and search for the global minimum of the ``fit-likelihood'' estimator
\begin{equation}
    \mathcal{S}(\delta) = 
    \frac{\chi^2}{\max_{\varepsilon,\sigma} \chi^2} +
    \frac{\Delta^2}{\max_{\varepsilon,\sigma} \Delta^2}
\label{eq:S}
\end{equation}
with the results displayed in Fig.~\ref{fig:chi2}.  
%
% ------------------------------------------------------------------------------- 
% ------------------------------------------------------------------------------- 
%
The global best fit values (including those for isotropic graphene) are
indicated with a star and their explicit values are given in
Table~\ref{tab:LJparams}.
\begin{table}[h]
\begin{center}
    \renewcommand{\arraystretch}{1.6}
  \begin{tabular}{  c  c  c  c  c }
   \hline\hline 
    $\delta$ & $0.00$ & $0.10$ & $0.25$ & $0.34$ \\ 
    \hline
    $\varepsilon\; [\si{\kelvin}]$ & $16.247(7)$ & $16.28(9)$ & $16.407(6)$ & $16.61(2)$ 
    \\ 
    $\sigma\; [\si{\angstrom}]$ & $2.739(7)$  & $2.782(8)$ &  $2.895(6)$ & $3.08(1)$
    \\ \hline \hline
  \end{tabular}
\end{center}
\caption{\label{tab:LJparams}The optimal values of the Lennard-Jones parameters
which best reproduce the long distance continuum van der Waals tail of the
adsorption potential for a helium atom above strained graphene. The uncertainty
in the final digit is indicated in parenthesis where the error can be
attributed to the starting position height $z_0$ of the residual $\chi^2$ and the
functional form of the cost-distance function $\Delta^2$}
\end{table}
Again there is flexibility in the specific form of the likelihood estimator  
$\mathcal{S}(\delta)$ in Eq.~\eqref{eq:S} and we have investigated the effects of
using other functions, including different weightings of $\varepsilon$ and
$\sigma$ as well as a relative scale factor between $\chi^2$ and $\Delta^2$.
These ambiguities add an additional source of error (along with the starting
$z$-coordinate of the residual) that is reflected in the
quantitative uncertainties reported in Table~\ref{tab:LJparams}. These errors,
which are on the order of a few percent, do not affect the observed qualitative
dependence on strain: both Lennard-Jones parameters are monotonically
increasing functions of $\delta$.  

We note, rather remarkably, that for isotropic graphene, we recover the
experimentally determined parameters $\varepsilon_0$ and $\sigma_0$ used for
helium interacting with a graphite surface \cite{Carlos:1980cf}.  This result
provides a novel and independent theoretical verification of the validity of
these parameters, as the inputs to our calculation only include the dynamical
polarizability of helium defined in Eq.~\eqref{eq:alpha} and the well known
properties of graphene in vacuum.

As strain is increased, both $\varepsilon$ and $\sigma$ grow, with $\sigma$
being most strongly affected, (increasing by over $10\%$ for $\delta=0.34$).
This is the expected behavior, as it encapsulates the geometric properties of
the potential and sets the distance at which the attractive minima occurs for a
two-body interaction.  $\varepsilon$, which sets the energy of the minimum,
increases by $2.5\%$ at the highest strain considered. This different response
to strain is likely indicative of their role in the potential,
Eq.~\eqref{eq:Usum}, where $\varepsilon$ sets a linear scale while $\sigma$
appears with the sixth power of the distance and thus has a greater effect on
the long distance tail. 

\subsection{Results: strained Lennard-Jones potential}
\begin{figure}[t]
\begin{center}
\includegraphics[width=1.0\columnwidth]{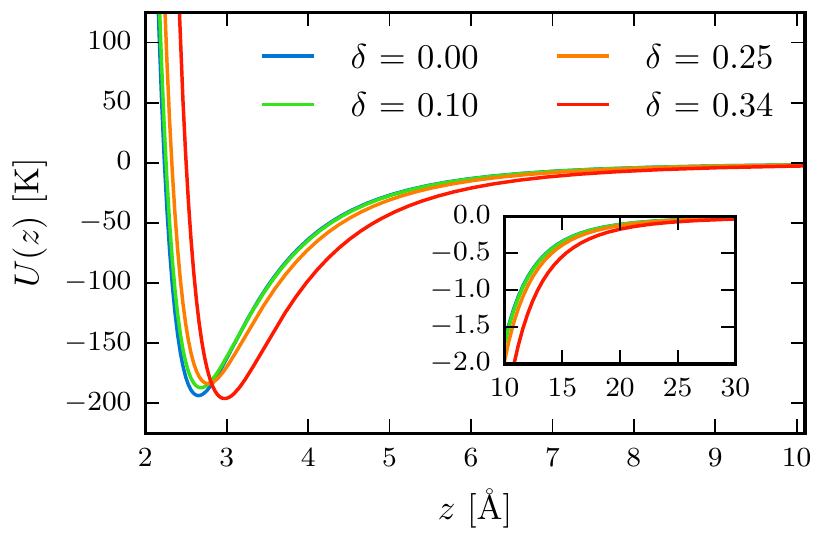}
\end{center}
\caption{(Color online) The Lennard-Jones adsorption potential for a He
adatom placed at coordinate $\vec{r}_A = (0,0,z)$ above a graphene sheet with
uniaxial strain along the armchair direction parameterized by $\delta$.
The inset (same axes) shows the long distance tail of the
potential, with increasing strain causesing the dispersion force to increase,
in agreement with continuum van der Waals calculations in the long distance
limit.}
\label{fig:Uz_strain}
\end{figure}

%-------------------
\label{sub:results}
Having determined the strain dependent Lennard-Jones parameters in
Table~\ref{tab:LJparams} we now compute the complete form of the
many-body adsorption potential for a He adatom above strained graphene,
with the results shown in Fig.~\ref{fig:Uz_strain}.
%
% ------------------------------------------------------------------------------- 
% ------------------------------------------------------------------------------- 
%
Here the helium atom is centered with respect to the hexagonal graphene unit
cell ($\vec{r}_A = (0,0,z)$, as in Fig.~\ref{fig:VzIsoNoFit}) and we observe
that the location of the attractive minima, $r_m$ is pushed to
larger distances above the sheet as the strain increases, with a concomitant
softening (increase) of the potential from
$U(r_m) \simeq \SI{-192}{\kelvin}$ for isotropic graphene with $\delta =
0$ to $U(r_m) \simeq \SI{-182}{\kelvin}$ at $\delta = 0.25$.  For the
strongest strain, $\delta=0.34$, we find
that the location of minima is pushed out to a distance of $r_m \simeq
\SI{2.95}{\angstrom}$, but in contrast to weaker strain, its depth decreases to
$U(r_m) \simeq \SI{-194}{\kelvin}$ indicating a propensity for
enhanced adsorption.  We believe that this behavior may be indicative of a
breakdown of our our fitting procedure at large strain as it only weights deviations in the
long distance tail and neglects the corrugated structure of the lattice at short
distances.  This is confirmed in the next section via ab-initio calculations.
During the fit, the large increase of the vdW force found in the
continuum approximation at large strain is most efficiently captured through an
increase in $\sigma$.  For two particles, changes in $\sigma$ only alter the
location of the potential minimum, whereas the maximum depth of the many-body
adsorption potential is strongly dependent on this hard-core radius as well as
the relative coordination between the adatom at the graphene lattice as seen in
Fig.~\ref{fig:VzIsoNoFit}.  $\varepsilon$, on the other hand, has the
same effect on both the two- and many-body potential, setting an overall linear
energy scale.

\begin{figure}[t]
\begin{center}
\includegraphics[width=1.0\columnwidth]{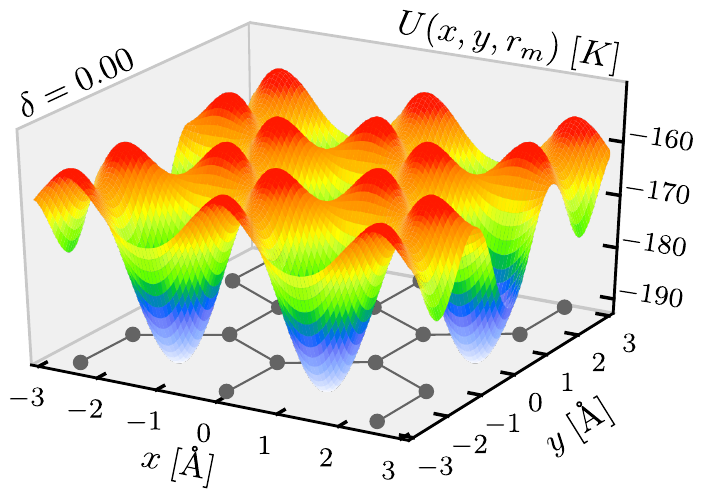}
\includegraphics[width=1.0\columnwidth]{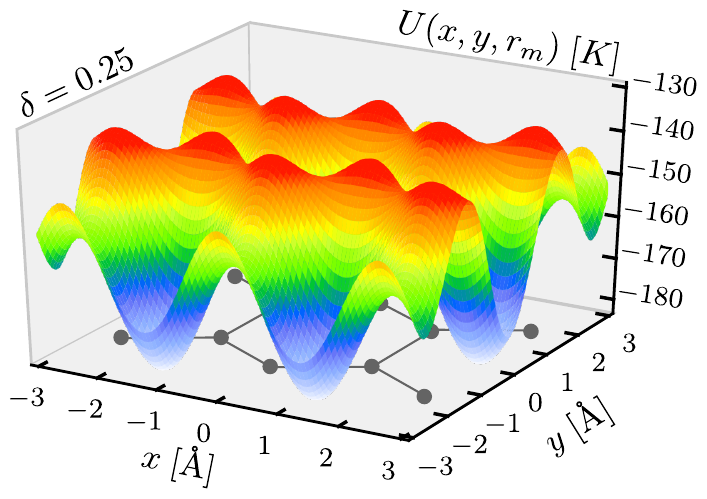}
\end{center}
\caption{(Color online) The spatial dependence of the Lennard-Jones adsorption potential for a
    He atom at a fixed distance $z=r_m(\delta)$ above an isotropic ($\delta=0$, top) and
strained ($\delta=0.25$, bottom) graphene sheet using the parameters in
Table~\ref{tab:LJparams}.}
\label{fig:Uxy3d}
\end{figure}

\begin{figure}[t]
\begin{center}
\includegraphics[width=1.0\columnwidth]{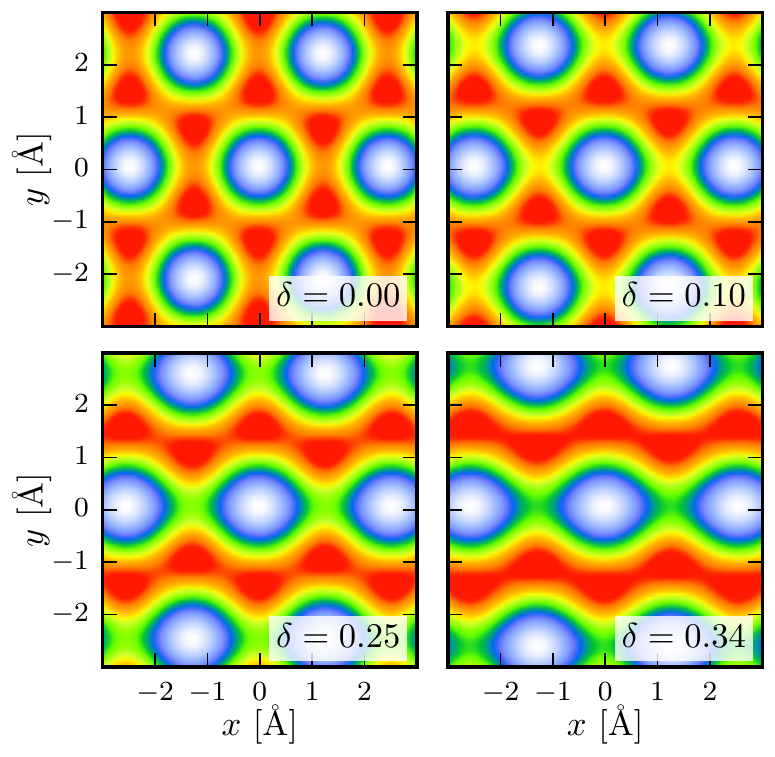}
\end{center}
\caption{(Color online) The Lennard-Jones potential $U(x,y,r_m)$ for different values of the
strain parameter $\delta$ for a helium adatom located at fixed $z=r_m$ above the
graphene sheet where $r_m(\delta) \simeq 2.635, 2.663, 2.768, 2.951\
\si{\angstrom}$ is the strain dependent position of the minimum for 
$\delta=0.0,0.1,0.25,0.34$ respectively. Each panel has been independently normalized
such that the color scale ranges from $\text{min}_{x,y}\ U(x,y,r_m)$ (white) to
$\text{max}_{x,y}\ U(x,y,r_m)$ (red).}
\label{fig:Uxyrm}
\end{figure}

To better understand these effects, we fix $z = r_m(\delta)$ and evaluate the
adsorption potential $U(x,y,r_m)$ as a function of the $x$ and $y$ coordinates
as seen in Fig.~\ref{fig:Uxy3d}. 
%
% ------------------------------------------------------------------------------- 
% ------------------------------------------------------------------------------- 
%
For unstrained graphene (top panel), we observe modulations on the order of
$15\%$ as the atom is moved laterally at fixed $z$.  The potential has an  
egg-carton structure with global
minima occurring at hexagon centers and giving rise to the 
$\sqrt{3} \times \sqrt{3}$ $\mathrm{R}30^{\circ}$ commensurate phase
experimentally observed in graphite \cite{Bretz:1971jo,Zimmerli:1992hz,
Dash:1994cp}.  This phase, where helium atoms occupy $1/3$ of the strong
binding sites, has also been observed for isotropic graphene in Monte Carlo
simulations \cite{Gordillo:2009jb, Kwon:2012ie, Happacher:2013ht}.  In the
presence of large strain, the potential is more washboard-like, with high
ridges tracking the zig-zag direction and deep minima, again centered at the
hexagon centers, but with a reduced energy barrier between them.  The evolution
of these coordination effects with strain are more apparent when normalizing
deviations of the potential between their minimum and maximum values as seen in
Fig.~\ref{fig:Uxyrm}, where again we have fixed $z=r_m(\delta)$.
%
% ------------------------------------------------------------------------------- 
% ------------------------------------------------------------------------------- 
%
The valley-to-peak difference in the potential increases from
approximately $\SI{36}{\kelvin}$ for $\delta=0$ to $\SI{49}{\kelvin}$ for
$\delta=0.25$ while the energy barriers between minima are systematically
reduced along the zig-zag troughs.

\begin{figure}[h]
\begin{center}
\includegraphics[width=1.0\columnwidth]{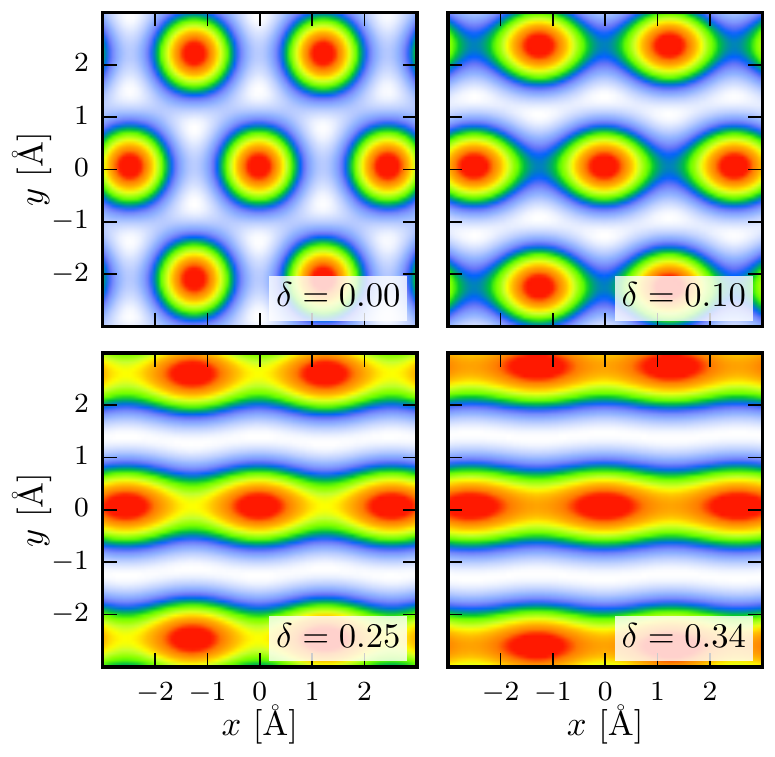}
\end{center}
\caption{The Lennard-Jones potential $U(x,y,2\sigma_0)$ for different values of the
    strain parameter $\delta$ for a helium adatom located at $z=2\sigma_0 =
\SI{5.48}{\angstrom}$ above the graphene sheet.}
\label{fig:Uxy2sigma}
\end{figure}

If we increase the fixed height above the sheet and set it to the strain
independent constant $z=2\sigma_0 = \SI{5.48}{\angstrom}$ we find very
different behavior as seen in Fig.~\ref{fig:Uxy2sigma}.
%
% ------------------------------------------------------------------------------- 
% ------------------------------------------------------------------------------- 
%
The location with respect to the lattice of peaks and valleys has now reversed,
with the hexagon center always representing the maxima in the potential.
While the variations in the potential are suppressed as $z$ increases: $\Delta
U(z=2\sigma_0,\delta=0.0) \simeq \SI{0.27}{\milli\kelvin}$ and 
$\Delta U(z=2\sigma_0,\delta=0.25) \simeq \SI{1.8}{\milli\kelvin}$,  the nearly $600\%$
increase demonstrates the large range of mechanical tunability of vdW
interactions in this system.  We note that the distance $z=2\sigma_0$
corresponds to the approximate location above the graphene sheet where 
a second layer of helium is adsorbed \cite{Gordillo:2009jb, Kwon:2012ie} whose
properties are still under debate \cite{Happacher:2013ht, Nakamura:2014vx}. 

In summary, we have found that in order to reproduce the increase in
the vdW attraction between a helium atom and a deformed graphene surface at
large distances computed within Lifshitz theory, it is necessary to employ
strain-dependent Lennard-Jones parameters.  At short distances, these modified
parameters in conjunction with the deformed lattice structure produce a highly
anisotropic, yet weakened adsorption potential with minima pushed to higher
energies at a location further above the graphene compared to the unstrained
case.  For the largest strains we considered ($35\%$), the potential minima is
pushed to nearly $\SI{3}{\angstrom}$ above the substrate. However, in contrast
to weaker strains, the depth of the potential well is slightly greater than
that for isotropic graphene.  A these extreme deformations, there is some
ambiguity in the relationship between the velocity anisotropy $v_y/v_x$ and the
strain percentage $\delta$ which requires an extrapolation procedure. This
uncertainty in combination  with a reduction in confidence of our fitting
method in this high-strain regime, indicates that a closer look at the short
distance potential is warranted.  This can be accomplished via a first
principles determination of the dispersion energy.

\subsection{\emph{Ab initio} calculations for coronene}
%------------------------------------------------------
\label{sub:ab_initio_calculations_for_coronene}
In this section, we calculate from {\em ab initio} methods the interaction
potential of a single He atom situated at a distance $z$ above the center of an
aromatic molecule, which represents a reasonable model for the near-field
interaction of the atom with a graphene plane.  The interaction of neutral
atoms and molecules with graphene is dominated by dispersion terms, leading to
van der Waals-type potentials, as discussed above.  The {\em ab initio}
evaluation of dispersion terms is delicate, requiring accurate treatment of the
correlation energy \cite{Cramer,Bartlett}. Two methods are considered reliable
enough for this determination \cite{Cramer}: M\o{}ller-Plesset \cite{MP} or
coupled cluster \cite{CC}.  Whereas the latter is considered of higher
precision, its computational cost is prohibitive for larger molecular clusters
and provides relatively small quantitative gains. Thus, we have performed 
calculations using the 2nd order M\o{}ller-Plesset perturbative approach which
captures about 95\% of the correlation energy \cite{Cramer}.

\begin{figure}[t]
\begin{center}
    \includegraphics[width=1.0\columnwidth]{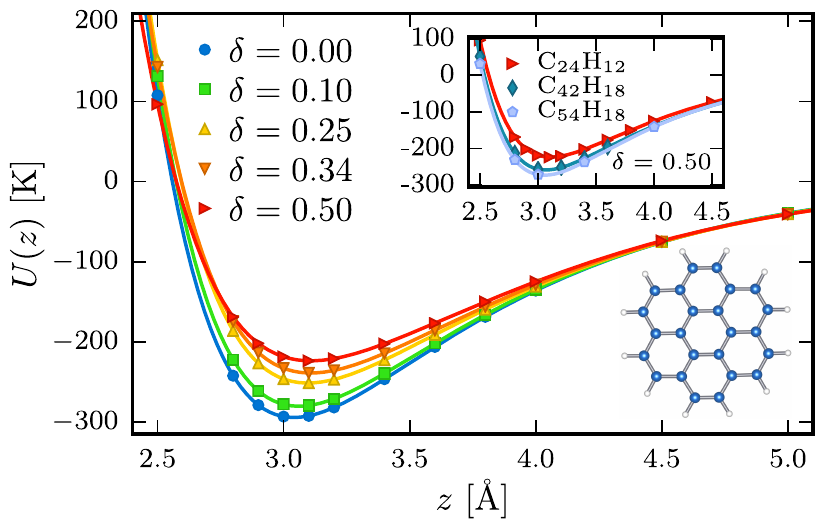}
\end{center}
\caption{(Color online) Adsorption potential for helium above (centered) a single strained
    coronene (C$_{24}$H$_{12}$) molecule (lower right) calculated in the 2nd order M\o{}ller-Plesset
\cite{MP} approximation using a 6-31++G(d,p) basis set \cite{Pople}.
Upper left: dependence of the adsorption potential on molecular size calculated
for $\delta = 0.50$ in coronene (C$_{24}$H$_{12}$)  hexabenzocoronene
(C$_{42}$H$_{18}$), and circumcoronene (C$_{54}$H$_{18}$) (same axes as main
panel). Similar
size-dependence is observed for lower strains.}  
\label{fig:rUstrain}
\end{figure}

All calculations
were performed in Gaussian 09 \cite{g09} using the Pople-type \cite{Pople}
6-31++G(d,p) basis set which includes diffusion of all orbitals and polarization
functions d for carbon and p for helium. For the aromatic molecules
representing graphene, we utilized coronene (C$_{24}$H$_{12}$, lower inset,
Fig.~\ref{fig:rUstrain}) or strained coronene, with the carbon atoms situated at
positions given in Eqs.~\eqref{eq:latticeVectors}--\eqref{eq:basisVectors},
i.e., no geometry optimizations were performed on the aromatic carbons which
would have eliminated the strain (the positions of the hydrogen terminators
were optimized in each configuration). The energy of the system was computed
for various values of the distance $z$ between the He atom and the aromatic
plane, and the asymptotic energy for $z \rightarrow \infty$ was removed as a
baseline (obtained by extrapolation of the energies for $z$ = 10, 15, 20, and
30 {\AA}).

The results for the interaction
potential of He on strained coronene are shown in Fig.~\ref{fig:rUstrain}. The upper inset shows the dependence of
this potential on the size of the aromatic compound. We find, in agreement
with the calculations of Section~\ref{sub:results}, that strain has two
dominant and connected effects on the helium adsorption potential: the
potential minima is pushed outwards from the sheet (as compared to isotropic
molecules) causing the attraction strength to be diminished.  Within our first
principles numerical calculations, this trend is monotonic with increasing
strain, further supporting the hypothesis that the previously employed fitting
procedure breaks down for highly deformed graphene lattices.  The absolute
value of the energy of the adsorption potential minima differs substantially
between Figs.~\ref{fig:Uz_strain} and \ref{fig:rUstrain} due to the presence of
hydrogen terminators necessary for chemical stability.

%
% ------------------------------------------------------------------------------- 
% ------------------------------------------------------------------------------- 
%

\section{Conclusions and Outlook}
\label{sec:conclusions}

In conclusion, we have analyzed in detail the van der Waals potential of  three
atoms (He, H, and Na) with uniaxially strained graphene ranging from
weak to moderately strong.  While these atoms have very different static
polarizabilities (Na being the most polarizable and He the least) and
characteristic frequencies, leading to very different potential strengths, the
overall dependence of their van der Waals potential on graphene strain is quite similar.
The potential is sensitive to strain and always increases, which can be traced
back to the enhanced graphene polarization. Since the enhanced polarization
also leads to increased screening of the Coulomb potential, as 
described by Eqs.~\eqref{e2}, \eqref{eq:Vz1} and \eqref{C3rel},  the exact value 
of the  van der Waals potential  increase reflects the delicate balance between higher polarization
and screening.  Our calculations show that  enhancement of the van der Waals
potential can be as high as $100\%$ for strong strain $\delta \approx 35\%$.
For such large values we always keep in mind that the strain is in the
armchair direction to ensure that the system remains semi-metallic (i.e. in the
anisotropic Dirac fermion ``universality class"). While it is unrealistic to
expect that graphene itself can be used in this extreme regime, the development of
artificial anisotropic graphene-like lattices as well the continuous stream of
discoveries in the field of 2D atomic crystals  could provide a potentially
exciting and fruitful playground for the phenomena we describe in this paper.
As mentioned in Section~\ref{sec:intro}, 
exampes of such anisotropic  systems include   graphene superlattices
\cite{chpark,chpark2,rusponi,herb}, tunable honeycomb optical lattices
\cite{tarruell}, and  molecular graphene  \cite{gomes}.  
Additional systems of interest
 could include atomically thin ${\mathrm{MoS}}_{2}$ \cite{Heinz-MOS, Maria}
which exhibits  strain-sensitive band structure \cite{MOS-1, MOS-2, MOS-3}, as
well as graphene on hexagonal boron nitride (h-BN), with a superlattice of
spontaneous strain fields and strong electron correlation effects \cite{HBN-1,
HBN-2, HBN-3}. Analysis of these and other 2D materials requires extensions of
the present work in several directions, such as inclusion of spectral gap,
spin-orbit coupling, and gauge fields induced by more complicated strain
configurations,  among others \cite{Maria}.

We have applied our results on the strain dependence of the van der Waals
interaction to the problem of quantum reflection, finding that it can be
significantly suppressed by strain. Pragmatically, this implies that cold atoms
on strained  graphene-based lattices  can approach the surface and thus
experience strong inelastic scattering (usually accompanied by emission of
flexural phonons in the substrate).  In this regime, dissipative many-body
phenomena \cite{Dennis} become of great importance as strain is applied; these
are by themselves complex theoretical  problems which we leave for future
studies.

Finally, we have explored the effects of mechanical strain on the
helium-graphene adsorption potential near the surface, finding that it can be
drastically modified.  By matching the results of long-distance continuum
calculations of the van der Waals interaction with an effective sum over
two-body interactions for He above a strained graphene lattice, we have
independently determined phenomenological Lennard-Jones parameters for the
system, finding agreement with common values used for the helium-carbon
interaction.  As strain is increased, the parameters $\varepsilon$ and $\sigma$
for the two-body interaction grow monotonically.  While this causes an increase
in attraction far from the sheet, the strength of the resulting many-body
adsorption potential for helium near the surface is reduced.  The resulting
locations of potential minima reflect the anisotropy of the deformed lattice
and are pushed to larger distances above the sheet, causing weaker adsorption
with increased strain. This trend was confirmed via \emph{ab initio} calculations of a
single helium atom above aromatic nanographene molecules.

Mechanically tuning the helium-graphene adsorption potential presents a
fundamentally new approach to the problem of engineering novel low dimensional
liquid phases, providing a method to inhibit classical wetting and promote
collective behavior. The formation of connected adsorption potential valleys in
Fig.~\ref{fig:Uxyrm} for 25\% strain may allow for adatoms to minimize
their kinetic energy by spatially delocalizing along them, offering a
mechanism that may favor anisotropic first layer superfluidity at low
temperature.  At smaller (and more experimentally realistic) values of strain,
the first layer may remain commensurate, but the second adsorbed layer, which
should be both anisotropic and weakly bound, would be an ideal candidate to
form a two dimensional quantum liquid.  This possibility is particularly
exciting in light of the fact that the exact nature of the second
layer of helium adsorbed on a \emph{graphite} surface is still under debate
\cite{Greywall:1993eg, Shibayama:2009jn} with recent heat capacity measurements
indicating the possibility of an exotic quantum hexatic state
\cite{Nakamura:2014vx}. The introduction of a mechanical strain into the
arsenal of experimental tuning parameters may help to uncover and confirm the
existence of this and other predicted quantum liquid phases.

\section{acknowledgments}

We are grateful to Dennis Clougherty for numerous stimulating discussions
related to the subject of this work and we acknowledge M. Cole for his insights
into the adsorption of helium on graphite.  The research of V. N. Kotov was supported
by the U.S. Department of Energy (DOE) grant DE-FG02-08ER46512.

\bibliographystyle{apsrev4-1}
\bibliography{refs}

\end{document}